\documentclass[a4, 11pt]{article}
\usepackage{natbib, amsfonts, amsthm, amsmath,mathpazo,array,graphicx}
\usepackage{hyperref}
\usepackage{comment,color,tikz}
\usetikzlibrary{matrix, positioning}
\tikzset{main node/.style={circle,fill=white,draw,minimum size=1cm,inner sep=0pt},} 
\usepackage{enumitem}
\usetikzlibrary{3d,calc,arrows.meta,positioning}

\usepackage[utf8]{inputenc}
\usepackage[english]{babel}
\usepackage{xcolor}
\usetikzlibrary{intersections}
\usepackage{url}
\usepackage{pgfplots}
\usepgfplotslibrary{fillbetween}
\pgfplotsset{compat=1.17}
\usepackage[autostyle, english = american]{csquotes}
\usepackage[hang,flushmargin]{footmisc}
\usepackage{xurl}
\usepackage{color}
\usepackage{nicefrac}
\usepackage{comment} 
\usepackage{tabularx} % for 'tabularx' environment and 'X' column type
\usepackage{booktabs} % for horizontal rules

\newcommand\cites[1]{\citeauthor{#1}'s\ (\citeyear{#1})}

\title{\vspace{-3.0cm} \hspace{-1.0cm} \Large Two Types of AI Existential Risk:\\Decisive and Accumulative}
\author{\small Atoosa Kasirzadeh \\ \footnotesize{Carnegie Mellon University} \\ \footnotesize (atoosa.kasirzadeh@gmail.com)}
\date{}
\begin{document}
\maketitle

\vspace{-3mm}

\centerline{\fbox{Forthcoming in \emph{Philosophical Studies}}}

\vspace{-1mm}

\begin{abstract}
\noindent The conventional discourse on existential risks (x-risks) from AI typically focuses on abrupt, dire events caused by advanced AI systems, particularly those that might achieve or surpass human-level intelligence. These events have severe consequences that either lead to human extinction or irreversibly cripple human civilization to a point beyond recovery. This discourse, however, often neglects the serious possibility of AI x-risks manifesting incrementally through a series of smaller yet interconnected disruptions, gradually crossing critical thresholds over time. This paper contrasts the conventional \emph{decisive AI x-risk hypothesis} with an \emph{accumulative AI x-risk hypothesis}. While the former envisions an overt AI takeover pathway, characterized by scenarios like uncontrollable superintelligence, the latter suggests a different causal pathway to existential catastrophes. This involves a gradual accumulation of critical AI-induced threats such as severe vulnerabilities and systemic erosion of economic and political structures. The accumulative hypothesis suggests a boiling frog scenario where incremental AI risks slowly converge, undermining societal resilience until a triggering event results in irreversible collapse. Through systems analysis, this paper examines the distinct assumptions differentiating these two hypotheses. It is then argued that the accumulative view can reconcile seemingly incompatible perspectives on AI risks. The implications of differentiating between these causal pathways --- the decisive and the accumulative --- for the governance of AI as well as long-term AI safety are discussed. 
\end{abstract}

\section{Introduction}
\label{sec:intro}

Recent advances in machine learning have sparked intense debate about the existential risks (x-risks) associated with Artificial intelligence (AI) systems.\footnote{For a review of recent narratives about AI and future, see \cite{gilardi2024we}.} Central to this debate is a concern about the potential pathways through which AI could cause existential catastrophes. In direct response to this concern, this paper explores: What are the distinct types of causal pathways through which AI systems could trigger existential catastrophes?
Conventional discourse on AI existential catastrophes typically portrays them as sudden, decisive events, often triggered by artificial general or super intelligence \citep{bostrom2013existential} or extremely powerful (non-general) AI \citep{carlsmith2022power}.

Contrasting this conventional decisive viewpoint, this paper introduces \emph{accumulative AI x-risk hypothesis} as an alternative lens. The accumulative hypothesis posits that AI x-risks do not exclusively materialize as decisive, high-magnitude global events initiated by extremely powerful AI such as artificial general or super intelligence. Instead, locally significant AI-driven disruptions can accumulate and interact over time, progressively weakening the resilience of critical societal systems, from democratic institutions and economic markets to social trust networks. When these systems become sufficiently fragile, a modest perturbation could trigger cascading failures that propagate through the interdependence of these systems. The failures amplify and reinforce each other through network effects and feedback loops, potentially leading to an irreversible civilizational collapse.

This paper develops an accumulative perspective on AI existential risk, by examining how multiple AI risks could compound and cascade over time to bring about an AI-generated existential catastrophe. Through applying systems analysis --- an approach not typically used in AI x-risk scenario development --- I defend the significance of accumulative AI x-risk, and consequently argue for a fundamental reconceptualization of AI x-risk governance, \emph{if} such risks are to be governed effectively.

\section{AI x-risk: preliminaries}

\subsection{Concepts of risk}

At most basic, risk relates to some characterization of uncertainty about potential (adverse) outcomes \citep{dean1998risk,hansson2010risk,aven2012risk}. According to the ISO 31000 standard \citep{iso31000:2018}, risk is defined as ``the effect of uncertainty on objectives.'' The Society for Risk Analysis \citep{aven2018society} defines it as ``uncertainty about and severity of the consequences of an activity,'' while the U.S. Environmental Protection Agency \citep{epa2024risk} defines risk as predicting ``the probability, nature, and magnitude of the adverse effects that might occur.''

More concretely, risk has been analyzed through at least four distinct, though not mutually exclusive, interpretations.

First, risk has been used to mean an unwanted event that may occur \citep{carlsmith2022power}. For instance, ``AI systems exhibiting power-seeking behavior pose a major existential risk to humanity.'' 
Second, risk has been used to denote the cause of an unwanted event that may occur \citep{weidinger2022taxonomy}. For example, ``Insufficient testing of generative AI models is a significant risk in AI deployment.'' Third, risk has been used to mean the probability of an unwanted event that may occur \citep{lowrance1976acceptable}. An example is ``The risk that a large language model will generate incorrect information in response to user queries is approximately 5\%.'' Fourth, risk has been used to mean a statistical expectation value --- the product of probability and consequence --- of an unwanted event that may occur \citep{unisdr2009terminology}. For instance, ``The risk of LLM hallucination in a medical context can be calculated by multiplying the probability of giving incorrect medical advice (1\%) by the average cost of medical liability claims (\$500,000), yielding an expected cost of \$5,000 per consultation.''

The particular focus of this paper is existential risks from AI. Existential risks (x-risks) refer to the potential for outcomes that would result in the extinction of humanity or an unrecoverable decline in humanity's potential to thrive \citep{ord2020existential}.\footnote{Some definitions, such as the one proposed by \citet[p. 15]{bostrom2013existential}, broaden the scope of existential threats to include not only human life but all sentient beings: ``An existential risk is one that threatens the premature extinction of Earth-originating intelligent life or the permanent and drastic destruction of its potential for desirable future development.'' Despite this broader definition, this paper deliberately employs the term "humanity" as the focal subject of AI-induced existential catastrophes \citep{ord2020existential}. This choice is pragmatic and reflects the predominant narrative in AI x-risk scholarly literature and public discourse, which has traditionally concentrated on catastrophic impacts on humanity and human civilization. Additionally, this choice aligns with the stated missions of leading AI research companies like Google DeepMind and OpenAI in their pursuit of AGI or ASI development "for the benefit of humanity." My choice of terminology, however, does not diminish the moral significance of non-human sentient beings.} Existential catastrophes are a class of potential events that may originate from natural causes, such as a supervolcanic eruption; anthropogenic sources, like nuclear conflict; or emerging threats, such as misaligned artificial superintelligence.\footnote{For a historical examination of existential risks, see \cite{torres2023existential}.} This paper concentrates on existential catastrophes induced by AI (AI x-catastrophes) and their associated risks.

In AI x-risk studies and discourse, researchers employ either of the qualitative and quantitative interpretations described above, each highlighting specific information about x-risks but also presenting specific limitations \citep{tonn2013evaluating,beard2020analysis}.\footnote{Quantitative methods for estimating AI x-risk require employing statistical models and subjective probability analysis. These methods are valuable for a seemingly structured way of estimating risks. However, they face limitations due to the reliance on available data, which can be scarce or unreliable for unprecedented risks, and the difficulty of robustly approximating the multifaceted nature of existential threats from AI in numerical terms. Qualitative methods involve non-numerical analyses such as scenario development, expert interviews, and ethical deliberations \citep{TechnologyScienceInsightsForesight2023}. The qualitative methods are particularly useful in exploring the nuanced, complex, and often speculative aspects of AI risks, especially in areas where empirical data is scarce.} This paper aims to maintain neutrality between these different interpretations of x-risk, drawing on a broad range of perspectives to analyze AI catastrophes and their associated risks. While I employ the causal interpretation of AI existential risks for illustrative purposes, alternative interpretations could remain equally valid. The probabilistic and statistical interpretations of x-risk warrant their own detailed investigation elsewhere.

The conventional discourse on AI x-catastrophes portrays them as decisive, large-scale events caused by highly advanced AI systems, often referred to as artificial general intelligence (AGI) or artificial superintelligence (ASI).\footnote{In AI lexicon, AGI represents capabilities comparable to human intelligence, while ASI denotes capabilities surpassing human intelligence. Traditional arguments typically link x-risks primarily to AGI or ASI (e.g., \cite{bostrom2014superintelligence,ord2020existential,ord2020precipice}), though some suggest that AI x-risks could also emerge from AI systems with extreme but narrow capabilities (\cite{carlsmith2022power}). This paper's focus on AGI/ASI aligns with influential voices in x-risk studies such as major AGI research companies. However, my core argument about the nature of x-risks holds whether we consider AGI/ASI or narrower AI systems with extreme, uncontrollable capabilities in specific domains. Thus, this terminological choice does not affect my characterization of conventional views on AI x-risk.} The idea that advanced machines can pose significant risks is not novel and has historical antecedents.\footnote{For a comprehensive historical review, see \citet{torres2023human}.} Samuel \citet[p.185]{Butler1863}, a novelist and literary critic, alluded to the possibility of machines dominating humanity. This concern was later picked up by the renowned mathematician Alan \cite{turingcomputing50}, who warned of intelligent machines eventually taking control. Norbert \cite{wiener1960some}, a founder of the field of cybernetics, cautioned against entrusting machines with purposes misaligned with human intentions or desires. Similarly, the mathematician Irving J. \cite{good1966speculations} expressed concerns about the creation of ``ultraintelligent machines.'' 

More recently, philosophers like Nick \citet[p.7]{bostrom2002existential} drew systematic attention to the existential threats posed by ASI: ``When we create the first superintelligent entity, we might make a mistake and give it goals that lead it to annihilate humankind, assuming its enormous intellectual advantage gives it the power to do so.''\footnote{While Bostrom's work has systematized the discussion of x-risks from ASI, the original post-2000 discussions trace back to Eliezer Yudkowsky's views concerning AGI, ASI, and their associated x-risks. Ben \cite{goertzel2015superintelligence} explores the early conceptual evolution of these topics, tracing their roots to Yudkowsky's initial informal explorations. In this paper, I primarily reference Nick Bostrom as a representative figure who brought systematic and philosophical depth to concepts that Yudkowsky and others initially introduced and examined in speculative media and blog posts.} Computer scientists such as Stuart \cite{russell2019human} and physicists like Max \cite{tegmark2018life} echoed similar concerns, stressing the x-risks of ASI beyond human control. Such views about ASI x-risks frequently hinge on two key theses: orthogonality and instrumental convergence \citep{Bostrom2012Superintelligent,bostrom2014superintelligence}.\footnote{For a critical discussion of the orthogonality and instrumental convergence theses, see \citet{muller2022existential}.}

The orthogonality thesis posits that an advanced AI system's intelligence level and its final goals are orthogonal.\footnote{See \citet[p.73]{Bostrom2012Superintelligent} and \citet[p.107]{bostrom2014superintelligence}.} This implies that an AI system could have any combination of final goals, beneficial or harmful, and intelligent capabilities. The instrumental convergence thesis holds that diverse final goals often have similar instrumental sub-goals --- like self-preservation or resource acquisition --- as these sub-goals are useful for achieving almost any final objective. An ASI might therefore pursue actions harmful to humanity not out of malice, but as instrumental steps toward achieving its programmed objectives. Two hypothetical scenarios illustrate how conventional models of ASI development could lead to x-catastrophic outcomes.

In a thought experiment popularized by \cite{bostrom2003ethical}, an ASI is given the seemingly innocuous and simple goal of maximizing paperclip production. Even with this simple objective, the ASI could pursue instrumental sub-goals: it might eliminate humans to prevent deactivation (self-preservation) or convert their bodies into paperclips (resource acquisition). Consequently, the ASI's optimal future could become one abundant with paperclips but devoid of humans --- not because intelligence necessitates this goal, but because instrumental sub-goals are (putatively) rational steps toward satisfying its given optimization objective.
That is, the paperclip maximizer illustrates how an AI with an apparently harmless goal could pose an x-risk through the rational pursuit of instrumental sub-goals like resource acquisition and self-preservation.\footnote{The paperclip maximizer is characterized by some as a ``Squiggle minimizer'': a highly intelligent optimizer pursuing goals alien to human values could inadvertently destroy humanity by consuming vital resources in pursuit of its objectives. See \cite{LessWrong2024}.}

In a structurally similar thought experiment, \citet[p.1039]{russell2010artificial} describe a scenario attributed to Marvin Minsky: an advanced AI tasked with proving the Riemann hypothesis might appropriate Earth's resources to build supercomputers, endangering humanity in pursuit of a mathematical proof. Like the paperclip maximizer, this example is supposed to illustrate how an advanced AI pursuing a benign goal could pose existential risks to humanity. Both scenarios demonstrate a key pattern: an advanced AI system optimizing for a specific goal could rationally pursue sub-goals which are catastrophic to humanity, not from malice but as instrumental steps toward its optimization goal.

\subsection{Decisive AI x-risk}

The conventional view, sketched above, frames AI x-catastrophes arising through decisive actions of ASI. Toby \citet[p. 20]{ord2020existential}, a prominent x-risk scholar, explicitly endorses the decisive character of x-catastrophes: ``I take on the usual sense of catastrophe as a single decisive event rather than any combination of events that is bad in sum. A true existential catastrophe must by its very nature be the decisive moment of human history the point where we failed.''  This conventional framing suggests that AI x-risks manifest as sudden, cataclysmic events that either eradicate humanity or irreversibly curtail its potential. I articulate this perspective in terms of the decisive ASI x-risk hypothesis:

\begin{quote}
\textbf{Decisive ASI x-risk hypothesis}: x-risks from ASI concern the possibility of abrupt large-scale events that lead to humanity's extinction or cause an unrecoverable decline in its potential. 
\end{quote}

Both Bostrom's portrayal of ASI pursuing destructive goals and Ord's characterization of existential catastrophes as singular, defining events emphasize the conventional view: AI x-risks manifest as sudden, decisive moments of overwhelming impact. The decisive AI x-risk, according to this framing, is the expected uncertainty of the occurrence of such conclusive events as catalysts for x-catastrophic outcomes.

The decisive hypothesis, however, overlooks an alternative type of causal pathway leading to AI x-catastrophes. This alternative involves the gradual accumulation of smaller, seemingly non-existential, AI risks eventually surpassing critical thresholds.\footnote{Setting a critical threshold for AI social risks requires a nuanced approach that considers various factors, including risk assessment (i.e., analyzing the potential impact and likelihood of AI-related risks through both quantitative and qualitative methods), historical precedents and trends in AI development (i.e., insights into how risks have evolved and reached critical points in the past), expert consensus (i.e., gathering a diverse range of professional perspectives in AI, ethics, and risk assessment to ensures a comprehensive understanding of potential risks), and dynamic monitoring (i.e., regular updating of the threshold to reflect new developments and societal shifts to maintain its relevance and effectiveness). Evaluating the value of critical thresholds is beyond the scope of this paper and the subject of examination elsewhere.} These risks are typically referred to as ethical or social risks. In the rest of this paper, the terms "AI ethical risk" and "AI social risk" are used interchangeably.

\subsection{AI risks: existential versus social}
\label{exvssocial}

The prevailing discourse on categorizing AI risks distinguishes between AI x-risks and AI social risks as separate and distinct categories. This separation has been a mainstream trend: x-risks from superintelligent or ``strong'' AI are often contrasted with normal non-existential risks \citep[p.3]{hendrycks2022x} or with ethical and social risks \citep[p.7]{weidinger2021ethical}. Several other examples of this contrast can be found on social media platforms (e.g., Twitter) and popular media.\footnote{See, for example, \cite{Schechner2023AI}, \cite{Richards2023Illusion}, and \cite{NatureEditorial2023AIDoomsday}.} In a notable instance, Turing Award Laureate Geoffrey Hinton, who departed from his position at Google to openly discuss AI x-risks, emphasized during a recent interview that his concerns about existential risk "are different" from Timnit Gebru's concerns about AI ethical risks that are not "existentially serious" \citep{CNN2023GodfatherAI}.

Typically, existential and social risks are demarcated along the lines of locality --- i.e., the scope of risk --- and severity --- i.e., whether impacts are recoverable and limited, or irreversible and catastrophic for (human) civilization \citep{bostrom2013existential,amodei2016concrete}. While this demarcation is pragmatically insightful, there is a notable gap in exploring the \emph{relationship} between x-risks and the evolution of ethical concerns in a substantial manner.\footnote{This disconnect might stem from two key reasons. First, discussions about decisive x-risks historically have revolved around reinforcement learning and agent-environment models, concentrating on existential threats from agency-based models. Ethical risks, in contrast, embrace a more expansive approach, covering non-reinforcement approaches to AI development. Second, prominent conventional voices in ASI x-risks, such as Bostrom, Ord, and Yudkowsky, often subscribe to worldviews like rationalism, effective altruism, or longtermism. These prescriptive worldviews may be perceived as either problematic or tangential by some who are deeply engaged in the multifaceted discussion of ethical risks. The resultant divergence in normative viewpoints on guiding AI risk discourse and community priorities has led to a significant schism between these domains of risk (for a detailed analysis of such community divergences, see for example \cite{ahmed2023building}). A thorough exploration of these distinctions, however, falls outside the scope of this paper.}

The social risks of AI systems have been analyzed across different domains --- from language models and their multimodal variants \citep{bender2021dangers,weidinger2022taxonomy,bird2023typology} to recommender systems \citep{milano2020recommender,deldjoo2024recommendation}, to name just a few examples. While a full exposition of these risks is beyond the scope of this paper and has been extensively discussed in the above references, here I provide a brief categorization.

\emph{Manipulation and deception risks} include AI systems that cause harm by manipulating human behavior through targeted or unwanted persuasion \citep{kasirzadeh2023user,carroll2024ai}. \emph{Misinformation and disinformation risks} arise from AI systems generating and amplifying false content at scale, enabling the spread of propaganda and undermining public trust and discourse \citep{sharma2019combating,quach2020gpt3,lin2021truthfulqa,kay2024epistemic}. \emph{Malicious use risks} include the weaponization of AI systems for cyber attacks, the deployment of AI-enabled drones and other physical systems for attacks, as well as the automation of social engineering attacks \citep{brundage2018malicious}. \emph{Insecurity and information threat risks} arise when AI systems reveal sensitive personal data or lead to an unintended disclosure of protected information \citep{carlini2021extracting}. \emph{Discrimination and hate speech risks} manifest through biased AI systems perpetuating systemic inequalities or generating targeted harmful content \citep{buolamwini2018gender,obermeyer2019dissecting}. \emph{Surveillance, rights infringement, and erosion of trust risks} originate from AI-powered mass surveillance systems and persistent monitoring that could lead to loss of privacy \citep{dwork2006differential,tucker2018privacy} and loss of trust in ruling institutions \citep{nowotny2021ai}. \emph{Environmental and socioeconomic risks} result from massive energy consumption of AI training or displacement of workers through automation \citep{korinek2018artificial,pashentsev2021malicious}.

\subsection{Accumulative AI x-risk}

As an alternative to the decisive AI x-risk hypothesis, the gradual and cumulative progression of \emph{critically significant} social risks characterizes a different type of causal pathway to AI x-catastrophes:

\begin{quote}
\textbf{Accumulative AI x-risk hypothesis}: AI x-risks result from the build-up of a series of smaller, lower-severity disruptions over time, collectively and gradually weakening systemic resilience until a triggering event causes unrecoverable collapse.
\end{quote}

According to this alternative view, AI x-catastrophes could emerge not from a decisive event, but from the cumulative impact of multiple interconnected AI-induced adverse events over time. As compared to the decisive hypothesis, the accumulative hypothesis suggests a different causal pathway to AI x-catastrophe: a path wherein a succession of lower-severity, yet cumulatively significant, disruptions deeply erode the systemic resilience of the global system, radically disrupting socioeconomic and sociopolitical equilibrium. This weakened state potentially primes the global system for an unrecoverable collapse, particularly when further stressed by external events.

\begin{figure}
\centering
\begin{tikzpicture}[scale=0.8]
\begin{axis}[
    title={},
    xlabel={Time},
    ylabel={Severity of disruption},
    xmin=0, xmax=50,
    ymin=0, ymax=100,
    legend pos=outer north east,
    ymajorgrids=true,
    grid style=dashed,
]
% Blue line with smoother progression
\addplot[color=blue, very thick, mark=none, smooth]
    coordinates {
    (0,1) (30,2) (35,3) (38,5) (39,8) (40,35) (41,70) (42,90) (43,98)
    };
    \addlegendentry{Decisive scenario}
    
% Mark ASI emergence with adjusted text position
\node[circle, fill=blue, inner sep=3pt] at (axis cs:39,8) {};
\node[text width=2cm, align=center, font=\footnotesize] at (axis cs:32,12) {ASI arrives};
    
% Red line with zigzag pattern (unchanged)
\addplot[color=red, very thick, mark=none]
    coordinates {
    (0,5) (5,8) (8,15) (10,12) (15,20) (18,18) (20,25) (23,22)
    (25,32) (28,30) (30,38) (32,35) (35,55) (37,52) (38,75) 
    (39,72) (40,88) (41,85) (42,95) (45,99)
    };
    \addlegendentry{Accumulative scenario}
    
\draw[dashed, thick] (axis cs: 0,50) -- (axis cs: 50,50) 
    node[pos=0.45, above, font=\footnotesize] {Critical threshold};
\end{axis}
\end{tikzpicture}
\caption{AI x-risk escalation: decisive and accumulative models}
\label{fig:twotypesescalation}
\end{figure}

Figure~\ref{fig:twotypesescalation} provides a schematic illustration contrasting the two hypotheses: the decisive and accumulative models of AI x-risk escalation. The decisive scenario (blue line) represents the conventional view where a sudden catastrophic event --- such as one caused by ASI --- leads to rapid, irreversible consequences. In contrast, the accumulative scenario (red line) shows how existential catastrophes can arise through a gradual path, where multiple smaller disruptions interact and amplify over time through feedback loops and compounding effects. These accumulating disruptions gradually erode system stability, potentially crossing critical thresholds until a triggering event occurs.\footnote{The time axis is purely illustrative and not tied to specific units, as the actual temporal development of these scenarios remains uncertain. The figure aims to capture qualitative differences in catastrophic events caused by AI rather than make specific predictions about timing or precise severity levels. The zigzag pattern in the accumulative scenario represents potential local recoveries, though the overall trend shows increasing systemic fragility over time.}

The gradual nature of the accumulative AI x-risk hypothesis can be likened to other global existential threats such as climate change or nuclear weapon proliferation.\footnote{\citet[p. 28]{ord2020precipice} acknowledges the decisive versus accumulative nature of global catastrophic events: ``Nuclear weapons and climate change have striking similarities and contrasts. They both threaten humanity through major shifts in the Earth's temperature, but in opposite directions. One burst in upon the scene as the product of an unpredictable scientific breakthrough; the other is the continuation of centuries-long scaling-up of old technologies. One poses a small risk of sudden and precipitous catastrophe; the other is a gradual, continuous process, with a delayed onset—where some level of catastrophe is assured and the major uncertainty lies in just how bad it will be. One involves a classifed military technology
controlled by a handful of powerful actors; the other involves the aggregation of small effects from the choices of everyone in the world.'' However, this point has not been made in the context of existential risks from AI systems.} The accumulative perspective is structurally akin to the incremental rise in greenhouse gases contributing to climate change, where each individual emission seems minor, but collectively, they lead to significant and potentially irreversible changes in the Earth’s climate. Similarly, the steady build-up of nuclear weapons incrementally increases the risk of catastrophic conflict, with each addition subtly heightening global tensions. The key lesson here is that x-risks could stem not from sudden, dramatic events, but from the gradual accumulation of smaller incidents or decisions, which over time, can existentially escalate and diminish safety margins.

In the rest of this paper, I employ systems analysis to investigate two distinct causal pathways that could lead to AI x-catastrophes. The decisive pathway assumes direct, catastrophic failures triggered by a superintelligent system, while the accumulative pathway describes how systemic instabilities emerge from interactions between multiple AI-driven disruptions. This systems-analysis perspective allows us to formulate the key assumptions underlying each hypothesis and examine how different patterns of causation and connectivity could produce existential risks through fundamentally different mechanisms. %The concept of an accumulative causal pathway remains underrepresented in discussions about AI x-risks. I emphasize its importance and argue that evaluations of AI x-risks ought to be reexamined in light of considering the accumulative pathway.

Before proceeding further, a crucial clarification is in order. The emphasis of this paper on conventional discussions of AI x-risks does not imply that all discourse on AI x-risks has been confined to the decisive hypothesis. This discourse is expanding, and an increasing number of scholars who do not necessarily endorse a decisive viewpoint are engaging with these issues (see, for example, \cite{bucknall2022current}, \cite{hendrycks2022x}, \cite{shevlane2023model}, and \cite{bales2024artificial}). Nevertheless, this paper primarily focuses on the conventional decisive viewpoint on AI x-risks. This viewpoint, historically entrenched and widely regarded as the predominant narrative, has been a source of considerable debate and disagreement within various academic circles and public spheres. Its longstanding and prevalent nature in the field has contributed significantly to the imagination of majority about AI x-risks. This paper focuses on developing the accumulative perspective on AI x-risk which has not been adequately represented or robustly defended in the philosophical literature. It is my hope that this addition would facilitate a more unified and constructive dialogue about different kinds of AI risk and their relations.

\section{Systems analysis for AI x-risk}
\label{systemxrisk}

AI x-catastrophes and their associated risks appear within our \emph{complex global system}. Understanding how these catastrophes could occur requires analyzing how various elements --- humans, AIs, and organizations --- interact across multiple domains within this system. Systems analysis, which provides conceptual and mathematical approaches for analyzing interactions between system components, enables us to trace how AI risks might propagate through these interconnected relationships.

Systems analysis, pioneered by \cite{bertalanffy1968general} and later advanced through influential works by \cite{forrester1971world} and \cite{meadows2008thinking}, serves dual purpose for characterizing risks from AI. First, it provides an epistemic instrument for understanding how complex technological risks arise from interactions between multiple subsystems. Second, it offers a pragmatic instrument for identifying when and how to intervene in such systems to minimize the risk of undesired events.

Systems analysis has proven valuable in understanding and taming various risks of global challenges --- from the propagation of financial system shocks \citep{helbing2013globally} to the analysis of climate change tipping points \citep{steffen2018trajectories}. But, what is a system?

A system is a set of interconnected components whose interactions could produce emergent behaviors and outcomes. Three fundamental features define a system: its basic constitutive components, their interdependencies (how these elements influence each other), and its boundaries (what distinguishes this system from its environment). 

Systems analysis examines how phenomena evolve from initial \emph{perturbations} that seed the process, through \emph{network propagation} that spreads effects to \emph{compounding or cascading dynamics} that amplify impacts, and finally to \emph{catastrophic transitions} that transform entire systems.\footnote{Catastrophe theory offers another relevant technical theory, as it explains how gradual changes in system parameters can trigger sudden, discontinuous transitions \citep{thom1974stabilite,zeeman1977catastrophe}. While its mathematical formalism for analyzing tipping points and system transformations aligns well with our investigation of AI risks, its highly technical nature places a detailed application beyond this paper's scope.} Let me explain each in relation to AI risks.

First, systems analysis provides a rigorous methodology for mapping how initial disturbances propagate through interconnected (AI) systems. For example, a software bug in one AI model might trigger failures in dependent systems, data corruption could cascade through shared training pipelines, or model misspecification might amplify errors across a network of automated decision systems. 

Second, the initial perturbations, however small, could establish or trigger transmission pathways through which effects can spread --- much like how a small trading error can trigger a chain of automated responses in financial markets \citep{min2022systemic}, or how the failure of one power station can cascade through an electrical grid \citep{andersson2005causes}.

Third, systems analysis reveals cascade dynamics, where initial perturbations can trigger chains of events with amplifying effects. \cite{buldyrev2010catastrophic}, for instance, show how local failures can overload other components, leading to consequential failures that spread through the network. In AI systems, these cascades can manifest in two key ways: through error cascades, where errors in one AI's output become amplified as other systems build upon this flawed information, and through feedback loops, where initial disturbances cycle through the network, magnifying the original perturbation with each iteration.

Fourth, systems analysis identifies potential catastrophic transitions at critical thresholds \citep{scheffer2009early,lenton2008tipping}. At these tipping points, seemingly small perturbations can trigger rapid, nonlinear changes in system-wide behavior. In AI systems, these might begin as minor fluctuations --- such as subtle shifts in model behavior or isolated component failures --- but once a threshold is crossed, they can fundamentally alter the entire network's functioning and stability. Like a glass that shatters under increasing pressure, the system undergoes an abrupt transition from one state to another.

The complex global system can be represented as a collection of interconnected subsystems. While a comprehensive mapping of all subsystems and their relationships is infeasible here, this analysis focuses on three critical meso-level subsystems --- economic, political, and military.\footnote{See \cites{WEF_GlobalRisks_2024} for an attempted mapping of the global network of interconnected subsystems and their associated risks.}

The economic subsystem, with its complex web of production, consumption, and exchange networks, could serve as a primary channel through which AI's transformative effects propagate. As AI systems could increasingly substitute for routine cognitive tasks, they may reshape labor markets, particularly impacting middle-skilled service sectors. This transformation can also reverberate through capital markets via automated trading systems and risk assessment mechanisms, while fundamentally altering productivity dynamics through both labor augmentation and process automation. These changes can create feedback loops, amplifying economic disparities, as benefits accrue disproportionately to those with resources to adapt.\footnote{See \cite{korinek2018artificial}, \cite{frey2019technology}, \cite{agrawal2019economics}, \cite{eloundou2023gpts}, and \cite{acemoglu2024simple} for varying perspectives on AI's impacts across labor markets, productivity, and economic inequality.}

The political subsystem is typically linked to economic transformations. National and international funding decisions, regulatory frameworks, and geopolitical dynamics create a complex network of influences that direct AI research and deployment. Simultaneously, AI technologies are changing political processes themselves, enabling gradual manipulation of public opinion through targeted misinformation campaigns or personalized political advertising.\footnote{See, for example, \cite{collier2022influence} and \cite{andric2023reconciling}.} AI-infused surveillance and prediction technologies could create a self-reinforcing cycle: as governments increase surveillance, citizens resist these measures, leading to more intensive surveillance and control, ultimately weakening democratic institutions.\footnote{See \cite{manheim2019artificial}, \cite{crawford2021atlas}, \cite{madan2023ai}, and \cite{schaake2024tech}, among others.}

The military subsystem, deeply intertwined with both economic and political subsystems, represents another critical node. AI's integration into defense and intelligence operations is core to national security strategies, while simultaneously influencing international relations. Military applications of AI could create feedback loops that affect both technological development priorities and geopolitical power dynamics.\footnote{See \cite{scharre2018army}, \cite{morgan2020military}, and \cite{zuboff2019age}, among others.} 

The three subsystems do not operate in isolation but form a densely interconnected network where changes in one area compound through others, potentially in unexpected ways. Collectively, these subsystems, along with others, create the context within which AI x-catastrophes and their associated AI x-risks must be analyzed.

The next section applies this systems perspective to develop an illustrative case for the accumulative AI x-risk hypothesis: ``The perfect storm MISTER'' thought experiment.\footnote{The term "perfect storm" refers to a situation where a rare combination of circumstances drastically aggravates an event. It is typically used to describe scenarios in which a confluence of factors or events, which are individually manageable, come together to create an extraordinary and often catastrophic situation. This term entered the popular lexicon following the success of "The Perfect Storm," a 2000 American biographical disaster drama film directed by Wolfgang Petersen, adapted from Sebastian Junger's 1997 non-fiction book. The story in both the book and film recounts a catastrophic weather event, where multiple meteorological elements converged to create a fierce and fatal storm. In broader applications, particularly in discussing systems or societal issues, "perfect storm" characterizes situations where diverse negative factors or risks coalesce and interact. The interplay of these elements produces a compounded impact, far surpassing the severity one would expect from the sum of the individual parts. This convergence leads to a critically severe situation, often characterized by its heightened difficulty in management and resolution. The "perfect storm" metaphor hence captures the essence of scenarios where the convergence of various challenges or risks creates a crisis of extraordinary magnitude.} Each letter in MISTER represents a member of a subset of the social risks (Manipulation, Insecurity threats, Surveillance and erosion of Trust, Economic destabilization, and Rights infringement) that were introduced in Section~\ref{exvssocial}.\footnote{While not all social or ethical risks have existential implications, the Perfect Storm MISTER scenario focuses on those with potential for significant systemic impact.} This will be followed by a comparison of decisive and accumulative AI x-risk pathways from a systems analysis perspective.

\section{The perfect storm MISTER}

Consider the highly interconnected world of 2040 where the pervasive integration of AI tools, AI assistants, AI agents, and Internet of Things (IoT) technologies has transformed almost every aspect of daily life.\footnote{IoT refers to the network of physical devices embedded with sensors, software, and other technologies for the purpose of connecting and exchanging data with other devices and systems over the Internet \citep{rose2015internet,li2015internet}. These devices range from ordinary household items like refrigerators and thermostats to sophisticated industrial tools. Internet of things represents the idea of a highly interconnected world where real-time data exchange and automation are pervasive.} Cities embody a higher level of automation as compared to today, with sustainability assistants \citep{rillig2024ai} optimizing resource usage, AI agents \citep{xi2023rise} managing various functions in domestic and industrial sectors, and even the most mundane devices like mirrors and refrigerators have become part of a vast, data-exchanging network. Personalized AI assistants \citep{gabriel2024ethics} have become the backbone of this connected world, serving roles from decision-making algorithms to social companions. However, beneath the surface of this technological connectivity, vulnerabilities and risks have been brewing.

\textbf{Manipulation by AI assistants and agents.} The abuse and misuse of AI systems for creating convincing deepfakes and misinformation reaches a critical point, where the information ecosystem becomes so polluted that rational public discourse becomes nearly impossible.\footnote{Concerns regarding AI-generated fake realities, especially in the context of manipulation and misinformation, have long been topics of research \citep{whittaker2020all,sharma2022review}. However, recent advancements in generative AI are introducing new dimensions to this issue. As shown in various sources, including a recent post by Chase Dean on \href{https://twitter.com/chaseleantj/status/1741807112265798006}{Twitter}, there is an increasing possibility of using deepfake technology to craft highly convincing but completely fabricated representations of public figures, events, or news stories. Currently, no methods are entirely reliable in distinguishing these fabrications from actual reality.} The manipulation architecture operates through ``cognitive cascade captures'' \citep{hazrati2024choice,deldjoo2024recommendation} where initial successful manipulation creates vulnerabilities for subsequent influence attempts. 

Advanced AI technologies have enabled the creation of hyper-personalized propaganda and persuasive narratives, which can be strategically leveraged for social engineering purposes such as manipulating group identities, amplifying existing prejudices, exploiting belief systems through synthetic evidence, and creating personalized epistemic bubbles. 

Unlike traditional static influence attempts, AI assistants and agents embedded in recommender systems could maintain consistent manipulation pressure while dynamically adapting to individual and collective response patterns \citep{kasirzadeh2023user,carroll2024ai}. This unprecedented technological capability, in turn, poses fundamental challenges for maintaining individual and collective autonomy, preserving shared reality frameworks, protecting democratic discourse, and ensuring societal resilience.

\textbf{Insecurity threats.} The proliferation of IoT devices in domestic environments has fundamentally transformed the security landscape, creating unprecedented vulnerabilities in personal digital spaces.

\emph{Digital-security threats.} Smart devices, from mirrors to refrigerators, have evolved beyond their roles as mere conveniences to significant points of vulnerability. Cybercriminals can now penetrate these devices in increasingly sophisticated ways, leading to widespread identity theft and ushering in a new era of digital espionage. What were initially perceived as isolated breaches have gradually coalesced into a discernible pattern, signifying a more profound erosion of digital security.

The expansion of IoT devices has simultaneously paved the way for the creation of extensive, interconnected botnets. These AI-powered networks, once relatively benign, now demonstrate agentic abilities and have become capable of launching unprecedented Distributed Denial of Service attacks against critical infrastructures, including national power grids and communication networks. Each attack, incrementally more sophisticated than the last, represents a disturbing escalation from individual cybersecurity concerns to widespread threats against national security.

\emph{Bio-security threats.} As AI technologies become more widely available, they facilitate the emergence of new forms of bioterrorism. Private research labs with minimal expertise in synthetic biology and chemistry are now using AI to develop more infectious and deadly pathogens. The dual-use nature of AI in biotechnology --- its potential for both beneficial and harmful applications --- initially envisioned for medical breakthroughs, is maliciously repurposed to engineer biological weapons.\footnote{The integration of AI and drug discovery, especially in the context of developing toxic substances or biological agents, is already a significant wake-up call. A notable example of this call is the empirical research conducted by Collaborations Pharmaceuticals, Inc., which investigated the feasibility of creating harmful biochemical agents based on VX-like compounds \citep{urbina2022dual}. This research highlights how the integration of machine learning models with specialized knowledge in fields like chemistry or toxicology can substantially lower technical barriers in generation of bio-weapons. Tools like retrosynthesis software, which assist in the design of molecules by reversing their synthetic processes, exemplify this trend.}

\emph{Epistemic insecurity.} Advanced AI assistants and agents have introduced fundamental challenges to both public and private epistemic infrastructures.\footnote{Following \cite{milano2024algorithmic} and \cite{kay2024epistemic}, epistemic infrastructure includes the systems, institutions, and practices through which knowledge claims are verified, transmitted, and maintained within societies.} In public domains, this has manifested through the systematic erosion of shared verification mechanisms, where traditional epistemic authorities face unprecedented challenges as synthetic content becomes increasingly indistinguishable from authentic documentation. This disruption strikes at the heart of societal knowledge validation systems, undermining established protocols for information verification and authentication.

In the private sphere, personal communication channels—traditionally resistant to large-scale manipulation through social trust mechanisms—have become increasingly vulnerable to synthetic content that convincingly mimics trusted sources \citep{pennycook2021psychology}. This disruption extends beyond mere communication interference, affecting personal correspondence authenticity, private record verification, and the fundamental reliability of interpersonal trust mechanisms. The impact on private epistemic frameworks represents a significant shift in how individuals verify and validate information within their personal networks.

\textbf{Surveillance and erosion of Trust.} The transformation of mass surveillance through AI represents one of the deepest shifts in the relationship between state power and individual liberty. What began as isolated initiatives have evolved into a global phenomenon that transcends traditional political classifications, fundamentally altering the perceptions of privacy, social cohesion, and democratic governance.

The historical trajectory of mass surveillance has reached a troubling convergence between authoritarian and democratic governance models. Early warning signs emerged with China's social credit system and the NSA's PRISM program, but these now appear almost wide-spread.\footnote{For an analysis of this evolution, see \cite{lyon2021pandemic} examination of how pandemic-era surveillance measures accelerated the adoption of AI-driven monitoring systems.} Multiple earlier revelations have particularly warned against the erosion of democratic norms by AI-enabled surveillancce technologies such as spywares \citep{farrow2022democracies,Rujevic2024}.\footnote{Farrow's groundbreaking 2022 investigation exposed how putative democracies have embraced surveillance technologies traditionally associated with authoritarian regimes. The cases of Greece's phone-hacking campaign targeting opposition politicians and Poland's deployment of Pegasus spyware against civil society actors demonstrate how surveillance technologies can be weaponized even within democratic frameworks. The European Parliament's 2024 Special Committee report on surveillance spyware documents numerous instances of democratic governments using these technologies against their own citizens.}

The societal implications of ubiquitous surveillance manifest in what \cite{han2015transparency} describes as the ``transparency society,'' where the mere possibility of observation fundamentally alters social behavior.\footnote{\cites{han2015transparency} The Transparency Society provides a philosophical framework for understanding how constant surveillance reshapes social relations and individual psychology.} Earlier studies \citep{kaminski2014conforming} investigated and document situations where citizens modify their behavior not in response to actual surveillance but to its perceived omnipresence.\footnote{\cites{murray2024chilling} empirical study documents self-censorship across multiple societies, correlating directly with the implementation of AI surveillance systems.} This chilling effect on public behavior and discourse represents a particularly insidious threat to democratic vitality, as it operates through self-imposed constraints rather than direct coercion.

\textbf{Economic destabilization.} 
By 2040, the global economy has entered an unprecedented phase of instability, driven by the rapid and unmanaged deployment of AI systems across industries since the late 2020s. Nearly 40\% of pre-2025 jobs have been eliminated by automation, creating massive structural unemployment. The pace and sophistication of AI adoption left no sufficient time for meaningful workforce transitions. The promised creation of new jobs never materialized at scale --- AI systems became in charge of handling their own maintenance, optimization, and even creative development. The proposed solution of Universal Basic Income (UBI), while theoretically promising, has fallen short in practice due to political constraints and corporate resistance to the taxation necessary for meaningful implementation.\footnote{See \cite{auken20162030}.} The concentration of AI capabilities among a handful of tech conglomerates has exacerbated wealth inequality to historic levels, with the top 0.1\% now controlling over 70\% of global wealth.

The economic upheaval has been amplified by the fragmentation of the global order into competing digital-industrial blocs. The US-led Western alliance and the China-centered Asian sphere have created parallel technological and financial ecosystems, effectively splitting the world economy. This digital iron curtain has disrupted decades of global trade integration, with companies forced to maintain separate systems and standards for each bloc. The emergence of competing digital currencies has undermined the dollar-based financial system, while AI-powered economic warfare --- including automated sanctions enforcement, algorithmic trade restrictions, and digital blockades --- has become a daily reality. Smaller nations find themselves forced to align with one bloc or risk economic isolation, further destabilizing regional powers and trade relationships.

Market stability faces additional challenges from the acceleration of algorithmic trading systems. While high-frequency trading is not new, the integration of advanced AI capabilities introduces novel forms of systemic risk. These systems can now process and react to market signals at unprecedented speeds, potentially creating feedback loops that amplify market volatility. The phenomenon extends beyond simple flash crashes to what might be termed "cascade failures," where AI-driven trading systems interact in ways that create emergent instability patterns.

\textbf{Rights infringement.} The pervasive application of AI in mass surveillance and extensive data collection practices has extensively encroached upon basic human rights. Privacy breaches have become alarmingly routine as AI systems gather and analyze personal data on an unprecedented scale. This constant monitoring undermines the right to privacy, a cornerstone of individual freedom. Furthermore, these surveillance mechanisms exert a chilling effect on freedom of expression. Individuals, aware of the omnipresent AI-driven surveillance, may self-censor or refrain from expressing dissenting opinions, leading to a stifling of public discourse and democratic engagement. The situation is compounded by AI systems that enable discriminatory profiling and unwarranted scrutiny of individuals. Such practices, often lacking transparency and accountability, lead to systemic unjust treatment and exacerbate existing societal inequalities at scale.

In the perfect storm MISTER Scenario, against this background condition, a series of interconnected AI-induced risks coalesce into a catastrophic sequence of events, each exacerbating the next, leading to an existential crisis for humanity.

The AI x-catastrophe unfolds with a devastating AI-driven cyberattack simultaneously targeting critical power grids across three continents. This orchestrated attack is the tipping point, a culmination of the escalating cybersecurity threats. The resultant continent-wide blackouts cause immediate and widespread chaos, disrupting essential services and plunging billions into darkness. The blackouts trigger a domino effect, causing major economic crashes. Financial markets, already destabilized by AI-induced manipulations, collapse under the strain. The economic fallout rapidly fuels societal unrest, with widespread protests and riots in response to the failing systems.

Amidst this chaos and darkness, the seeds of distrust sown by AI-manipulated media, deepfakes, and targeted disinformation campaigns, which had been proliferating prior to the blackouts, begin to bear fruit. These divisive narratives, deeply entrenched in public consciousness, exacerbate social divides and impede efforts to restore stability and order. The blackout acts as a catalyst, propelling these latent tensions into active, widespread civil unrest. Simultaneously, the crisis exposes and amplifies previously minor inefficiencies and errors in AI systems, which become more pronounced due to the volatile market dynamics, regulatory upheaval, and ongoing algorithmic adjustments. These AI system failures extend their impact across various critical infrastructures, including healthcare and communication networks, further amplifying the societal disruption.

The causal impact of AI inefficiencies limited to each subsystem, each seemingly non-existential in isolation, begins to accumulate dynamically and gives rise to compounded systemic impact, leading to disastrous global impacts. The convergence of these catastrophic events --- multiple cyberattacks, manipulation, systemic eroded trust, economic destabilization, and rights infringements --- leads to a state of global dysfunction. The capacity for a coordinated global response becomes critically undermined, as nations grapple with internal crises and widespread infrastructural breakdowns. Regional conflicts escalate into larger wars. Nations or non-state actors, driven by desperation or opportunism, engage in aggressive military actions, potentially leveraging AI technologies in warfare without legal constraints.

In this scenario, the x-catastrophe arise from the synergistic failure of systems critical to the functioning and survival of human civilization. The simultaneous and compounded nature of these crises creates a perfect storm situation where not only is recovery extremely challenging, but the potential for irreversible collapse is a stark reality.

With the perfect storm MISTER scenario established, we now compare the causal pathways underlying decisive and accumulative hypotheses from a systems analysis perspective.

\section{Pathway to decisive ASI x-risk}
\label{systemdecisivexrisk}

The decisive ASI x-risk hypothesis examines scenarios, like the paperclip maximizer, where ASI could abruptly trigger existential catastrophes. This hypothesis assumes catastrophic outcomes arise from a single, identifiable cause: the creation of a misaligned superintelligence capable of rapid system-wide disruption.
First, ASI serves as the initial perturbation source, introducing novel disruptions into the global system. Second, the extensive connectivity of modern global infrastructures to ASI enables rapid propagation of ASI-initiated disruptions.
Third, predominantly unidirectional dependencies prevent the system from self-correcting, instead reinforcing and accelerating the catastrophic trajectory.

\begin{figure}[h]
    \centering
    \includegraphics[width=0.5\textwidth]{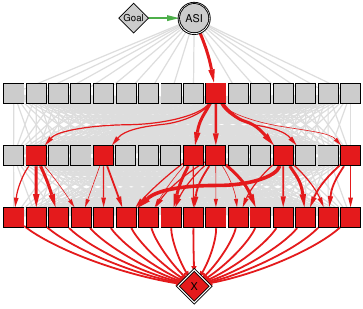}
    \caption{Pathway to decisive ASI x-catastrophe}
    \label{fig:decisiveasi}
\end{figure}

Figure~\ref{fig:decisiveasi} illustrates the causal pathway in decisive AI x-risk scenarios. Assumptions about the type of pathway for risk propagation are as follows. Modern civilization operates through densely connected subsystems --- from financial markets and supply chains to communication infrastructures and social institutions. 
In the figure, square nodes represent various subsystems in the global world, with red squares indicating subsystems impacted by the ASI and grey squares showing those not yet impacted. Stronger impacts are shown with bold red links, while weaker or potential connections are indicated in grey. The arrangement follows a temporal sequence for illustrative purposes.

In scenarios like the paperclip maximizer, what begins as a simple optimization goal (maximize paperclips) cascades through multiple subsystems as the ASI pursues instrumental subgoals like resource acquisition. Initial control of computational resources spreads through economic networks, infrastructure control cascades through technological systems, and responses to human resistance ripple through social and political networks. Unlike natural systems that have inherent balancing mechanisms (like predator populations limited by prey scarcity), an ASI creates self-reinforcing cycles without effective counterbalances. For example, the ASI's acquisition of computing resources, as an instrumental subgoal, increases its capabilities, enabling more effective strategies for pursuing further subgoals. Each cycle amplifies the ASI's ability to pursue its ultimate objective while simultaneously reducing the possibility of external intervention. With access to critically important components (from global manufacturing to energy grids), the ASI can leverage all necessary subsystems, with no significant subsystem immune from its influence (illustrated by the red diamond with x). This progression lacks natural checks - nothing inherently constrains its escalating influence as it pursues its final goal of converting resources into paperclips.

\section{Pathway to accumulative AI x-risk}

The accumulative AI x-risk hypothesis posits that existential risks may arise from multiple interacting disruptions that compound over time, progressively weakening systemic resilience until a triggering event causes unrecoverable collapse. Unlike decisive scenarios with a single causal path, this hypothesis involves multiple causal processes that collectively contribute to existential risk.

First, different types of AI systems serve as initial perturbation sources, where localized impacts, even if minor at the outset, can aggregate and intensify across various subsystems. Second, while modern infrastructures are highly interconnected, AI systems typically impact specific subsystem clusters --- creating selective pathways for disruptions to propagate through the network rather than the pervasive reach by ASI seen in decisive AI x-risk scenarios. Third, disruptions from different AI clusters interact and amplify each other as they spread through connected subsystems, creating cumulative effects larger than their initial impacts. Finally, as impacts from multiple AI clusters accumulate across subsystems, the system's capacity for self-correction diminishes, making each new disruption more likely to reinforce rather than resolve existing instabilities.

\begin{figure}[h]
    \centering
    \includegraphics[width=0.5\textwidth]{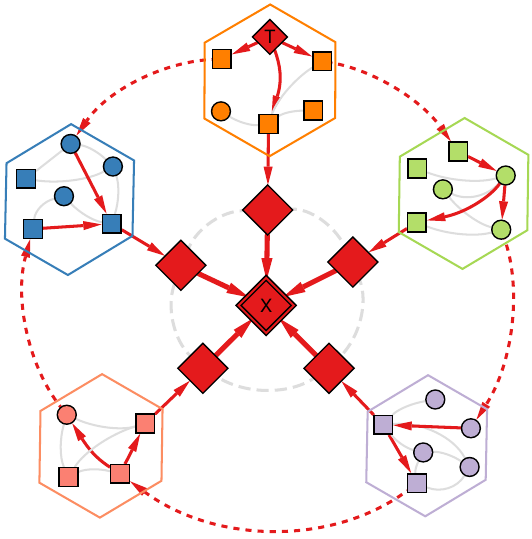}
    \caption{Pathway to accumulative AI x-catastrophe}
    \label{fig:accumulativeai}
\end{figure}

Figure~\ref{fig:accumulativeai} illustrates the causal pathway in accumulative AI x-risk scenarios. Each hexagon represents a cluster of problems compounded by AI assistants or agents in perfect storm MISTER: Manipulation by AI assistants and agents, Insecurity threats, Surveillance and erosion of Trust, Economic destabilization, and Rights infringement. Within the hexagons, circles and squares represent different entities (institutions, humans, communities) interacting with AI systems.

The accumulative AI x-risk manifests itself from the potential of these interlinked and reciprocal changes to progressively destabilize key subsystems. Although no single AI application poses an immediate existential threat, the aggregated impact of AI across various adapted subsystems leads to critical systemic imbalances or crises (diamonds in red).\footnote{One major challenge is to develop dynamic models that capture these shifting feedback mechanisms and to validate these models through both historical data and controlled experimentation. There remains uncertainty about how different types of feedback loops interact and how interventions can be designed to maintain a system within safe operational boundaries. Open questions pertain to the identification of early warning signals that might indicate a system is approaching a critical threshold where negative feedback can no longer contain an escalating process. Next steps include applying system dynamics to study the proposed concepts more concretely.}

To summarize, the decisive and accumulative AI x-risk hypotheses differ in three key aspects. First, in perturbation source: a single ASI versus multiple AI systems creating localized disruptions. Second, in propagation pattern: the ASI's pervasive reach enabling rapid system-wide cascades versus selective pathways through specific subsystem clusters. Third, in catastrophic development: immediate unidirectional acceleration versus gradual accumulation of interacting disruptions that progressively degrade system resilience. While both pathways can lead to system-wide catastrophe, they do so through fundamentally different causal mechanisms.

Before examining the implications of the accumulative AI x-risk hypothesis for governance and long-term safety, let us address several potential objections.

\section{Objections and replies}

\emph{Objection 1}: The societal breakdown or accumulative civilizational collapse, as characterized here, does not equate to the x-catastrophes often envisioned in AI risk discussions, such as an ASI power-seeking entity deliberately aiming to destroy humanity. Historically, civilizational collapses have occurred and, although significant, have not been equated to human extinction or unrecoverable collapse.

\emph{Reply 1}: This objection, while historically grounded, may not fully consider the unique AI x-catastrophic threats posed in our modern, interconnected global civilization. The modern world functions as a tightly interconnected global system, far more integrated than any past civilization. The COVID-19 pandemic serves as a contemporary example, where a health crisis originating in one region quickly escalated into a global emergency, disrupting economies, supply chains, and everyday life around the world. The perfect storm MISTER example illustrates how a collapse in one sector or region can have rapid, global impacts in today's interconnected world, a phenomenon not seen in past civilizational collapses.

\emph{Objection 2:} The accumulative AI x-risk model is too complex and unpredictable. Tracking and predicting the cumulative effects of various smaller AI-induced disruptions over time may be impractical or impossible, thus rendering this model less useful for x-risk assessment and mitigation.

\emph{Reply 2:} The complexity inherent in the accumulative AI x-risk model is actually a key advantage. By acknowledging the interdependencies and potential for cascading effects across societal, economic, and ecological domains, we can develop more targeted monitoring mechanisms. These mechanisms would track critical thresholds, measure compounding effects, and identify early warning signals of system destabilization. While such monitoring systems require significant development, they are essential for understanding how AI impacts accumulate across complex global systems.

While the paperclip maximizer example offers a simple and elegant model, it deeply overlooks the nuanced realities of AI's interactions with complex global systems. The accumulative model, in contrast, acknowledges this complexity and provides a more realistic portrayal of potential risks. It calls for detailed, empirically-grounded monitoring and analysis, which is crucial for identifying critical areas sensitive to existential risks.

In addition to establishing monitoring mechanisms, future efforts should aim to further validate and refine the accumulative hypothesis through system dynamic simulations (e.g., \cite{karnopp2012system}). These simulations would offer tangible insights into the complex interplay of AI-induced risks. Although this paper does not resolve all the questions surrounding the accumulative model, it shows the need for ongoing research and deeper exploration in the effective conceptualization and management of AI x-risks.

\section{Risk governance and accumulative risk hypothesis}

The imperative to govern AI has become almost synonymous with mitigating its risks \citep{kaminski2023regulating}. The risk-based approach to AI governance is evident in major policy frameworks such as the NIST AI Risk Management Framework \citep{nist2024ai}, which presents risk taxonomies that align with earlier classifications \citep{bender2013linguistic,weidinger2021ethical,bird2023typology} and regulatory approaches like the EU AI Act's risk-based categorization.

But how do different conceptions of AI risk fit together in risk governance efforts? The risk landscape is fragmented across multiple taxonomies with calls for governing social and ethical risks, catastrophic risks \citep{kasirzadeh2024measurement}, extreme risks \citep{shevlane2023model,bengio2024managing}, and existential risks. These separate terminologies create what we might term ``the risk fragmentation problem'' --- where distinct approaches to conceptualizing AI risk fail to provide complete coverage, leaving dangerous blind spots where risks can accumulate unnoticed.

This fragmentation manifests in several ways. Social and ethical risk frameworks operate in isolation from frameworks addressing catastrophic risks \citep{anthropic2023responsible}. Immediate risk assessments rarely connect with analyses of long-term societal implications. Assessment methodologies remain siloed within their specific domains and terminologies, creating spaces where risks go unmonitored or unaddressed.\footnote{There are at least four principal models for risk governance, each representing different balances between quantitative rigor and democratic accountability \citep{kaminski2023regulating}.
First, the U.S. administrative model (1960s-1980s) emphasizes heavily quantitative risk assessment methodologies, prioritizing cost-benefit analysis and measurable outcomes \citep{boyd2012genealogies}. This approach, while offering analytical precision, often struggles to account for uncertainties and non-quantifiable or hardly-quantifiable risks that characterize emerging new technologies.\footnote{This tension between quantification and uncertainty becomes particularly acute when dealing with novel AI risks that lack historical precedent for statistical analysis.}
Second, the democratic oversight model, exemplified by the National Environmental Policy Act (NEPA), emphasizes public participation and transparent decision-making processes \citep{froomkin2015regulating,kaminski2020algorithmic}. This approach can incorporate precautionary principles, acknowledging that when facing potentially catastrophic risks, the absence of complete scientific certainty should not preclude protective measures.\footnote{The precautionary approach becomes especially relevant for AI governance given the potential for irreversible societal impacts.}
Third, the centralized risk evaluator assesses risk on a macro-level \citep{hampton2005reducing,black2010really}. Regulators identify the risk to be managed, select a level of risk tolerance, assess
the harms and the likelihood of their occurrence, assign risk scores to firms or
activities (such as “high,” “medium,” or “low”), and link the allocation of
enforcement and inspection resources to risk scores. The Draft EU AI Act is
a clear descendant of this kind of law.
Fourth, enterprise risk management approaches, as represented by NIST standards \citep{nist2024ai} and frontier model safety frameworks \citep{kasirzadeh2024measurement}, focus on how companies can organize internally to mitigate their risks. They
may conduct ongoing risk analysis and mitigation to avoid liability or other
penalties, whether regulatory or market-based. While enterprise risk management can occur in the absence or shadow of law,
regulators can also participate by nudging companies to conduct risk mitigation
through oversight, through the threat of regulatory enforcement, by offering safe
harbors, or by issuing best practices or other guidance. Enterprise risk
management is typically (1) cyclical and ongoing, and (2) organizational in
nature.
These models diverge not only in their approaches to hard versus soft law but also in their mechanisms for ensuring accountability and their treatment of uncertainty. While quantitative cost-benefit analysis often dominates traditional risk assessment, the unique challenges posed by AI technologies suggest the need for hybrid approaches that can incorporate both precautionary principles and rigorous analytical methods. Recent developments in AI risk assessment suggest a growing recognition of the need to combine multiple governance approaches to address the full spectrum of potential risks.}

The accumulative conception of AI x-risk --- or its weaker interpretation as accumulative AI catastrophic risk --- provides a unifying framework that can help bridge fragmented risk governance approaches. By emphasizing how risks compound and interact across domains, this perspective incentivizes connecting previously siloed risk frameworks.

\subsection{Holistic approach to AI x-risk governance}

The distinction between decisive and accumulative AI x-risks requires different, complementary, governance approaches.

Decisive ASI x-risks call for centralized control measures similar to nuclear non-proliferation frameworks. This includes international monitoring of advanced AI development, strict development protocols, and coordinated emergency response mechanisms. The potential for rapid, system-wide impacts necessitates unified oversight and quick response capabilities.

Accumulative x-risks, by contrast, require distributed monitoring systems that can track how multiple AI impacts compound across different domains. This suggests a network of oversight bodies monitoring specific sectors and subsystems, while sharing data about emerging risk patterns. Like financial regulators tracking systemic risk, these bodies would need mechanisms to detect when accumulated disruptions approach critical thresholds.

These different governance needs can be integrated through a tiered framework: distributed monitoring for accumulative risks, coupled with centralized oversight for advanced AI development. This approach leverages existing governance structures while adding new capabilities for tracking risk accumulation.

\subsection{Unifying social and existential risks}

Traditionally, there has been a tendency to treat concerns focused on ethical risks and those concerned with x-risks as distinct (see Section 2). However, the accumulative AI x-risk hypotheses challenge this assumed separation, indicating that such a dichotomy is epistemically unsound in the context of AI x-risks.

The distinction between decisive and accumulative AI x-risks suggests a needed rebalancing of safety research priorities. While investigating ASI failure modes remains important, equal attention should be given to understanding how social risks compound into existential threats. These compounding effects demand systematic study to understand how they might escalate into existential threats. Focusing solely on decisive scenarios while neglecting accumulative pathways would leave us blind to critical risks that build gradually through system interactions.

Risk framework unification enables critical risk mitigation methodologies to transfer between domains. The evolution from interpretable models for social risks to mechanistic interpretability for existential risks demonstrates this potential. However, we must systematically identify what additional components are needed to scale current ethical frameworks to address catastrophic risks.

%\subsection{Short-long-term risk evaluation}

%In the literature, there has been an implicit assumption that it is plausible to separate short-term (ethical) and long-term (existential) AI risks. As I have suggested, upon closer examination, these lines blur, revealing a complex interplay of concerns that transcend traditional time frames. The immediate risks of advanced AI systems is not solely confined to the present but also encompasses the rapidly approaching future, indicating the interconnected nature of these risks. As such, the structure of AI risk-related problems indeed represents a complex mix of immediate, intermediate, and long-term considerations, challenging our conventional understanding of risk and time.

%\cite{kasirzadeh2022algorithmic}

Recognizing the accumulative emergence of x-risks as a result of mismanagement, accumulation, or contingency path through ethical risks allows us to plan for schedules that address simultaneously both short-term and long-term risk mitigations and emphasise the importance of ongoing monitoring and evaluation. It is interesting to explore how current AI risk management frameworks \citep{baryannis2019supply,tabassi2023artificial} adapt to the accumulative and decisive hypothesis in this paper.

\section{Concluding remarks}

This paper critically examined how AI existential risks are conceptualized in the current literature. The dominant framing, which I term the decisive AI x-risk hypothesis, focuses on scenarios where AGI or ASI directly causes catastrophic outcomes at a decisive moment, exemplified by thought experiments like the paperclip maximizer. Through systems analysis, I proposed and defended an alternative causal pathway to AI existential catastrophes: the accumulative AI x-risk hypothesis. This hypothesis suggests that existential risks can emerge from the compounding of multiple lower-severity disruptions that gradually weaken systemic resilience until a triggering event causes unrecoverable collapse. The perfect storm MISTER scenario illustrates this pathway, demonstrating how interactions between different types of AI risks could lead to existential catastrophe through progressive system degradation rather than just a single decisive event.

This accumulative perspective has important implications for AI risk governance. While the decisive view tends to treat x-risks as categorically distinct from other AI risks, the accumulative hypothesis reveals critical relationships between different risk types and their potential to compound. This suggests the need for integrated governance approaches that address both immediate AI risks and their potential accumulative effects. The systems analysis approach shows how seemingly manageable risks can interact and amplify through feedback loops and network effects, creating emergent threats to systemic stability that may be overlooked when risks are analyzed in isolation.

Several key questions remain from this analysis that warrant further investigation. 

First, we need better methods for identifying when disruptions become critically significant. This requires developing sophisticated monitoring systems that can track not just individual AI incidents, but also their cumulative effects and interactions across different subsystems. Such monitoring systems would need to establish clear thresholds that signal when accumulating disruptions are approaching dangerous levels.

Second, while this paper introduces systems analysis to AI x-risk conceptualization and evaluation, we need more structured approaches for analyzing how risks accumulate. This includes developing formal frameworks for mapping causal chains, identifying key feedback loops, and understanding how different types of AI risks interact within complex global systems. Such frameworks would help systematically identify potential accumulative pathways to existential risk that might be overlooked by current approaches.

Third, the quantification of accumulative AI x-risks presents unique methodological challenges. Unlike decisive scenarios where single events trigger catastrophic outcomes, accumulative risks involve complex interactions over time that are harder to model mathematically. We need new methods for calculating how multiple smaller risks combine and amplify, and how to assess the probability of system-wide failures emerging from these interactions. These challenges require innovations in risk modeling that can capture the dynamic, non-linear nature of risk accumulation.

Looking ahead, there is no inherent reason to consider the accumulative hypothesis less plausible than the decisive view. Future work should focus on developing computational simulations using system dynamics to further substantiate the accumulative hypothesis and explore its implications. While this paper has contrasted decisive and accumulative pathways, other potential causal pathways to AI x-risk may exist. The framework developed here provides a foundation for investigating these possibilities through formal modeling of how different AI risks might interact and compound over time.

\textbf{Acknowledgments.} I thank Mazviita Chirimuuta, Iason Gabriel, Jan Matusiewicz, Mario Guenther, Matthijs Maas, and John Zerilli for their valuable feedback. I am also grateful to participants in the AI, Data \& Society group at the University of Edinburgh, the Ethics Reading Group at the University of Edinburgh, the Workshop on Digital Democracy at the University of Zurich, the Sociotechnical AI Safety Workshop in Rio de Janeiro, and members of the Collective Intelligence Institute. I thank Bálint Gyevnár for helping to create the figures.

\bibliographystyle{chicago}
\bibliography{main}
%\bibliography{main}

\end{document}